\documentclass[12pt,preprint]{aastex}
\begin{document}

\def\fluxthres{\hat f_{\bar \e}}
\def\fluxeps{f_{_{\rm \epsilon}}}
\def\zmin{z_{\rm min}}
\def\zmax{z_{\rm max}}
\def\xmin{x_{\rm min}}
\def\xmax{x_{\rm max}}
\def\Estar{{\cal E}_*}
\def\Estarg{{\cal E}_{*\gamma}}
\def\Estargo{{\cal E}_{*\gamma 0}}
\def\Swift{\emph{Swift}}

\newcommand{\begeq}{\begin{equation}}
\newcommand{\fineq}{\end{equation}}
\newcommand{\begfig}{\begin{figure}}
\newcommand{\finfig}{\end{figure}}
\newcommand{\begeqarray}{\begin{eqnarray}}
\newcommand{\fineqarray}{\end{eqnarray}}
\newcommand{\e}{\epsilon}
\newcommand{\ep}{\epsilon^\prime}
\newcommand{\tp}{t^\prime}
\newcommand{\rb}{r_b^\prime}
\newcommand{\et}{\epsilon_T}
\newcommand{\dD}{\delta_{\rm D}}
\newcommand{\mup}{\mu^\prime}

%\slugcomment{to be submitted to \apj}

\shorttitle{GRB Predictions} \shortauthors{Dermer, Ramirez-Ruiz, \& Le}

\title{Correlation of Photon and Neutrino Fluxes in 
Blazars and Gamma Ray Bursts}
\shorttitle{Neutrinos from GRBs and Blazars}

\author{Charles D. Dermer\altaffilmark{1}, Enrico Ramirez-Ruiz\altaffilmark{2}, \& Truong Le\altaffilmark{1} }

\altaffiltext{1}{Code 7653, Naval Research Laboratory, Washington, DC 20375-5352; dermer@gamma.nrl.navy.mil, 
tle@ssd5.nrl.navy.mil}
\altaffiltext{2}{Department of Astronomy and Astrophysics, University of California, Santa Cruz, CA 95064; 
enrico@ucolick.org}

\begin{abstract}
Relativistic black-hole jet sources are leading candidates for high
energy ($\gg $ TeV) neutrino production. The relations defining (a)
efficient photopion losses of cosmic-ray protons on target photons and
(b) $\gamma\gamma$ opacity of $\gamma$ rays through that same target
photon field imply clear multiwavelength predictions for when and at
what energies blazars and GRBs should be most neutrino bright and
$\gamma$-ray dim. The use of multiwavelength observations to test 
the standard relativistic jet model for these source is illustrated.
\end{abstract}

\keywords{gamma-rays: bursts---neutrinos---theory }

\section{INTRODUCTION}

Multiwavelength observations provide important information about
radiation processes and properties of astrophysical sources that
cannot be obtained by observations in a narrow waveband.  The opening
of the high-energy neutrino window \citep[for review, see][]{lm00}
will provide a new channel of information that, in conjunction with
photon observations, will test models of these sources.  The km-scale
IceCube \citep{ahr04} reaches a total exposure of $\sim 1$ km$^3$-yr
in 2009 and its design sensitivity in 2011. Plans are also being made
for a northern hemisphere KM3NeT neutrino telescope in the
Mediterranean Sea.\footnote{For IceCube, see icecube.wisc.edu, and for
KM3NeT, see www.km3net.org.}

Because of their rapid variability, large luminosity, and relativistic
outflows, gamma-ray bursts (GRBs) and blazars have been considered as
the most likely sources of ultra-high energy cosmic rays and neutrinos
\citep{wb97,vie98,rm98,ahh00}.  The expected low neutrino-induced muon
event rate, even for the brightest $\gamma$-ray blazars
\citep{ad01,ns02} and GRBs \citep{da03,gue04}, and the increasing
importance of the cosmic-ray induced neutrino background at lower
energies \citep{kar03,ghs95}, means that event detection is greatly
improved by choosing appropriate time windows during periods of
highest neutrino luminosity.  The important time windows for neutrino
detection from blazars might be thought to be during $\gamma$-ray
flaring states and, for GRBs, around the $t_{90}$ time or during some
hours around the burst trigger, as suggested by the extended 100 MeV
-- GeV emission for GRB 940217 \citep{hur94} and delayed anomalous
$\gamma$-ray emission components in GRB 941017 and a few other GRBs
\citep{gon03,gon05}, because that is when energetic particle
acceleration is most vigorous.  But photopion production is enhanced
in conditions of high internal photon target density, so that times of
most favorable neutrino detection could also be argued to take place
during periods of low $\gamma$-ray flux as a result of attenuation by
the dense internal photon gas.

Here we explore this issue, considering how to use multiwavelength
observations (at optical, X-ray, and $\gamma$-ray energies for
blazars, and at X-ray and $\gamma$-ray energies for GRBs) to define
the most favorable conditions for efficient neutrino production.
Violation of these predictions will call into questions the underlying
assumptions currently used in models of GRBs and blazars.

\section{Analysis}

Blazars and GRBs are widely thought to be black-hole jet sources
powered ultimately by accretion onto a black hole, or by the spin
energy of the black hole. In both cases, observations show that
collimated outflows of highly relativistic plasma are ejected by
processes taking place from compact regions. In the internal shock
model \citep{mes06,pir05}, collisions between faster and slower shells
dissipate directed kinetic energy in the form of field energy and
accelerated particles that radiate.  After the collision, the
energized shocked fluid shell expands on the light-crossing time scale
$\rb/c$ or longer, where $\rb$ is the characteristic size of the
radiating fluid element in the comoving frame, assumed spherical and
isotropic in the comoving frame.  The causality constraint implies
that the size scale of the emitting region $\rb \lesssim c\dD
t_{var}/(1+z),$ where the
measured variability timescale $t_{var}= 10^\tau t_{\tau}$ s,
 and $\dD$ is the Doppler factor.

Within this geometry, the relation between the measured $\nu F_\nu$
flux $f_\e$ and the target photon emissivity
$j^\prime(\ep,\Omega^\prime) =d{\cal E}^\prime/dV^\prime dt^\prime
d\Omega^\prime d\ep$ is $f_\e \cong \dD^4 V_b^\prime \ep
j^\prime(\ep,\Omega^\prime)/d_L^2$, where $\e \equiv h\nu/m_e c^2$,
primes refer to comoving quantities, and unprimed quantities to
measured values.  The luminosity distance $d_L = 10^{\ell}d_{\ell}$ cm,
 can be calculated for the standard $\Lambda$CDM
universe ($h = 0.72$, $\Omega_m = 0.27$, $\Omega_\Lambda = 0.73$) for
a source at redshift $z$, $\ep j^\prime(\ep,\Omega^\prime) \cong c
u^\prime_{\epsilon^\prime}/4\pi\rb$ for radiation emitted
isotropically in the comoving frame, and $u^\prime_{\epsilon^\prime} =
m_ec^2 \e^{\prime 2} n^\prime(\ep )$ is the spectral energy density of
the radiation field.

We write the $\nu F_\nu$ flux as $f_\e = f_{\e_{pk}}S(x)$, where
$S(x)$ is a spectral function of the variable $x = \ep/\ep_{pk} =
\e/\e_{pk}$.  Here $\e_{pk}\equiv 10^j$ is the measured photon energy
(in units of $m_ec^2$) of the peak of the $\nu F_\nu$ spectrum with
peak flux $f_{\e_{pk}} = 10^\eta f_{\eta} $ ergs cm$^{-2}$ s$^{-1}$. 
Thus $\ep n^\prime(\ep) \cong
{3d_L^2 f_{\e_{pk}}S(x)/ (c r_b^{\prime 2} \dD^4 m_ec^2 \ep)}.$ The
rate at which protons lose energy through photohadronic processes is
$t_{\phi\pi}^{\prime-1} \cong \ep n^\prime(\ep)\hat\sigma c$, where
$\hat\sigma \cong 70~\mu$b is the product of the $\gamma p$
photohadronic cross section and inelasticity \citep{ad03}, and the
threshold condition $2\gamma_p^\prime \ep \gtrsim \ep_{thr} \cong 400$
relates the proper frame proton Lorentz factor $\gamma_p^\prime$ and
the internal photon energy.

The target comoving photon spectral energy distribution (SED) from
quasi-isotropic emissions, whether it is the synchrotron and
synchrotron self-Compton (SSC) fields or a cascade radiation field, is
approximated as a broken power law $S(x) = x^a H(1-x) + x^b H(x-1)$
with $\nu F_\nu$ indices $a$ and $b$ (see Fig.\ 1; more general
spectral forms can easily be treated). Here the Heaviside function
$H(x) = 1$ for $x>0$ and $H(x) = 0$ for $x <0$.  The photopion
energy-loss rate of ultrarelativistic protons with Lorentz factor
$\gamma_p^\prime$ interacting with photons with energy $\ep_{pk}$ near
the peak of the $\nu F_\nu$ spectrum is, from the proceeding
considerations,
\begin{equation}
\rho_{\phi\pi} =  {3\hat \sigma d_L^2 f_{\e_{pk}}
(1+z)\over m_ec^4 \dD^5 t_{var}^2 \e_{pk}}\;.
\label{rhophipi}
\end{equation}

For the model target photon spectrum with $0< a < 3$, $b < 0$,
\begin{equation}
t_{\phi\pi}^{\prime-1}(\gamma_p^\prime) \cong\rho_{\phi\pi}
\cases{ 2y^{b-1}/ [(1-b)(3-b)],  & $y \gg 1$  \cr
 2y^{a-1}/ [(1-a)(3-a)],  & $y\ll 1$, $0 < a \lesssim 1$ \cr
 (a-b)/ [(a-1)(1-b)],  & $y\ll 1$, $1 \lesssim a < 3\;$ \cr}\;
\label{tphipi}
\end{equation}
\citep{der07}, where $y \equiv \ep_{thr}/2\gamma_p^\prime 
\ep_{pk} \cong \dD^2 \ep_{thr}/2\gamma_p (1+z)\e_{pk},$
and the Lorentz factor $\gamma_p = E_p/m_pc^2$ of an escaping proton
as measured by a local observer is $\gamma_p \cong
\dD\gamma^\prime_p$.  The condition $y=1$ for protons with energy $E_p
= E_p^{\phi\pi}$ interacting with photons with energy $\e_{pk}$
implies that
\begin{equation}
E_p^{\phi\pi} \cong {m_pc^2 \dD^2\ep_{thr}\over 2(1+z) \e_{pk} }\cong 
 {1.9\times 10^{14} \dD^2 \over (1+z) \e_{pk}({\rm keV}) }\;{\rm eV}\;.
\label{Epk}
\end{equation}

The radiating fluid element will expand explosively following its
rapid energization through shell collisions, or through external
shocks formed when the outflow sweeps through the surrounding
medium. Photopion processes can be certain to be efficient---assuming
of course that ultrarelativistic protons are accelerated in black-hole
jets---if the photopion energy-loss rate $\rho_{\phi\pi}$, eq.\
(\ref{rhophipi}), is greater than the inverse of the light travel
timescale, $(1+z)/\dD t_{var}$. An energetic cosmic ray will therefore
lose a large fraction of its energy into electromagnetic and neutrino
radiations through photopion production when the jet Doppler factor
\begin{equation}
%\big( {3\hat \sigma \over m_ec^4 }\big)^{1/4}
\dD < \delta_{\phi\pi}  \equiv \;\big({3\hat\sigma d_L^2 
f_{\e_{pk}}\over m_ec^4 t_{var} \e_{pk}}\big)^{1/4} =
10^{-10.64+(2\ell + \eta - \tau - j)/4} 
%\cong 2.3\times 10^{-11}  10^{(2\ell + \eta - \tau - j)/4} 
\cong 7.3 d_{28}^{1/2}\;\big({f_{-10}\over t_0 \e_{pk}}\big)^{1/4}\;.
\label{Epk1}
\end{equation}

The same radiation field that functions as a target for photomeson
processes is a source of $\gamma\gamma$ opacity.  The photoabsorption
optical depth for a $\gamma$-ray photon with energy $\e_1$ in a
quasi-isotropic radiation field with spectral photon density
$n^\prime(\ep) $ is $\tau_{\gamma\gamma}(\ep_1)
\cong r_b^\prime \int_0^\infty d\ep \; \sigma_{\gamma\gamma }
(\ep,\ep_1)n^\prime(\ep ).$ Using a $\delta$-function approximation 
$\sigma_{\gamma\gamma}(\ep,\ep_1) \cong \sigma_{\rm T} \ep
\delta(\ep - 2/\ep_1)/3$ for the $\gamma\gamma$ pair production 
cross section $\sigma_{\gamma\gamma}$ 
\citep{zl85} gives 
\begin{equation}
\tau_{\gamma\gamma}(\e_1) \cong \tau_{\gamma\gamma}^{pk}\;
[({\e_1\over \e_1^{pk}})^{1-b}H(\e_1^{pk}-\e_1) + 
({\e_1\over \e_1^{pk}})^{1-a}H(\e_1-\e_1^{pk})]\;
\label{tauggap}
\end{equation}
\citep{der05} \citep[see also][]{ls01,bar06},
where
\begin{equation}
\tau_{\gamma\gamma}^{pk}= {\sigma_{\rm T} 
d_L^2 f_{\e_{pk}}\over 4 m_ec^4 t_{v} \dD^4 
\e_{pk}}\;
\label{tgge1}
\end{equation}
and
\begin{equation}
\e_1^{pk} = 
{2\dD^2\over (1+z)^2 \e_{pk}}\;.
\label{egge1}
\end{equation}
Fig.\ 1 compares the $\delta$-function approximation given by the
above equations with accurate calculations using the results of
\citet{gs67} \citep[see][for corrections]{bmg73}.

At the Doppler factor $\delta_{\rm D} = \delta_{\phi\pi} $ that allows
for efficient photopion production, the $\gamma\gamma$ optical depth
at photon energy $\e_1^{pk}$ is, after substituting eq.\ (\ref{Epk1})
into eq.\ (\ref{tgge1}),
\begin{equation}
\tau_{\gamma\gamma}^{\phi\pi} = {\sigma_{\rm T}\over 12 
\hat \sigma} \cong 800\;. 
\label{tauggpp}
\end{equation}
Whenever photopion production is important, $\gamma$ rays with
energies given by eq.\ (\ref{egge1}) have to be highly extincted by
$\gamma\gamma$ processes when interacting with peak target photons
with energy $\sim \e_{pk}$, making it impossible to detect $\gamma$
rays at these energies.  This energy is $$E_{\gamma}^{\gamma\gamma}=
{2m_ec^2\delta_{\phi\pi}^2\over (1+z)^2 \e_{pk}} = {2 m_e c^2
\;d_L\over (1+z)^2\e_{pk}^{3/2}}\;\sqrt{{3\hat \sigma f_{\e_{pk}}\over
m_ec^4 t_{var}}}
\cong
$$
\begin{equation}
{10^{-24.26 + \ell +(\eta-\tau -3j)/2}\;{\rm GeV}\over (1+z)^2}
 \cong 
{0.055 \;d_{28} f_{-10}^{1/2}\over (1+z)^2 t_0^{1/2}
 \e_{pk}^{3/2}}\;{\rm ~GeV}\;.
\label{epk2}
\end{equation}
The energy of protons that interact most strongly with peak target
photons through photopion processes under conditions when photopion
processes must be important is, from eq.\ (\ref{Epk}),
\begin{equation}
 E_p^{\phi\pi} = {m_pc^2 \delta_{\phi\pi}^2
 \e^\prime_{thr}\over 2 (1+z) \e_{pk}}
\cong 10^{-10 + \ell + (\eta -\tau -3j )/2}\;{\rm eV}\;
\cong 1.0\times 10^{13} d_{28} \; \sqrt{{f_{-10}\over 
t_0 \e_{pk}^3}}\;{\rm ~eV}\;.\label{eppk}
\end{equation}

\section{Results}

Table 1 lists the important quantities derived in this paper. The
quantity $\delta_{\phi\pi}$ is the jet Doppler factor where photopion
losses are guaranteed to be important for protons of escaping energy
$E_p^{\phi\pi}$. Protons with this energy undergo photopion
interactions primarily with peak target photons with energy $\sim
\e_{pk}$.  $E^{\gamma\gamma}_\gamma $ is the energy of $\gamma$ rays
that are attenuated through $\gamma\gamma$ pair production primarily
by peak target photons.

Consider the target photon variability time for the following source
classes: flat spectrum radio quasars (FSRQs) and GRBs, both known
sources of GeV radiation, and X-ray selected BL Lac objects (XBLs), of
which over a dozen are known TeV sources.  We define the variability
time scale $t_{var}$ as the measured time over which the absolute flux
varies by a factor of 2; if a source varies by $N$\% over time $\Delta
t$, then $t_{var} = 100\Delta t/N$, keeping in mind that this is a
conservative assumption for temporal variability given that quiescent
or unrelated emissions can add a separate slowly or nonvarying
background. R-band optical and RXTE PCA ($\approx 2$ -- 60 keV)
observations of 3C 279 and PKS 0528+134 show that day timescale optical and
X-ray variability can be expected for FSRQs \citep{har01,muk99}.  Day
timescale optical/UV and X-ray variability is also observed for TeV BL Lac objects
\citep{pia97,bla05}.

For canonical FSRQ values taken from observations of 3C 279 or PKS
0528+134, Table 1 shows that photopion production is already important
at Doppler factors of $\sim 9$ -- 16 during times of day-scale optical
flaring, and these optical photons effectively extinguish all $\gamma$
rays with energies $\gtrsim E^{\gamma\gamma}_\gamma/800^{1/(1-b)} $
(Fig.\ 1), which would certainly include all $\gtrsim 100$ GeV -- TeV
photons. Unfortunately, TeV telescopes have not so far been successful
in detecting FSRQs, but monitoring of an FSRQ during an optical flare
with low-energy threshold air Cherenkov telescope such as MAGIC would
identify periods of likely neutrino emission. Photohadronic neutrino
secondaries have energies $E_\nu\approx E_p^{\phi\pi}/20$ and so would
be produced at $\sim 10^{17}$ -- $10^{18}$ eV, providing possible
sources for ANITA \citep{bar06}, though outside IceCube's optimal energy
range.

For guaranteed importance of photohadronic production implied by
day-scale X-ray variability, the Doppler factors of FSRQs and TeV BL
Lac objects like Mrk 421 have to be unexpectedly small, $\lesssim
3$. If the X-ray flaring timescale of FSRQs were hourly rather than
daily, then $\delta_{\phi\pi}$ would more nearly correspond to Doppler
factors $\sim 5$ -- 10 inferred from unification studies and
superluminal motion observations of blazars \citep{vc94,up95}. During
episodes of highly variable X-ray flux, such sources should be
invisible to GLAST, and $\gg$ TeV neutrinos should be created if FSRQs
are sources of ultra-high energy cosmic rays. For the XBL estimate, a
15 minute X-ray flaring timescale has already been assumed. Thus FSRQs
are more likely than BL Lac objects to be high-energy neutrino sources
for IceCube, which also follows if, as is likely, the external
radiation field plays a strong role in neutrino production
\citep{bp99,ad01,ad03}.

The outcome of this analysis to identify periods of high-energy
neutrino production is best for bright GRBs with peak fluxes of
$\approx 10^{-6}$ ergs cm$^{-2}$ s$^{-1}$ and peak photon energy in
the range 50 keV -- 0.5 MeV that show $\lesssim 1$ s spikes of
emission. The bulk factors, $\approx 100$, are consistent with widely
considered outflow speeds in GRBs \citep[see, e.g.,][who also treat
$\gamma\gamma$ attenuation in GRBs]{rmz04}. Perhaps 100 MeV photons
could be observed, but the GLAST LAT should see no $\gtrsim $ GeV
photons if $\dD \lesssim \delta_{\phi\pi}$, which is the most
favorable time for detecting 100 TeV -- PeV neutrinos and is at an
optimal energy for detection with km-scale neutrino telescopes. Bright
X-ray flares with durations $\sim 10^2$ s observed hundreds to
thousands of seconds after the GRB trigger, like those discovered with
Swift \citep{obr06,bur07,mn06}, with blast wave Doppler factors
$\approx 50$, are also promising times to look for neutrinos and a
$\gamma$-ray spectrum attenuated above $\sim 100$ GeV $\gamma$ rays.

The condition $\tau_{\gamma\gamma}(\e_1)< 1$ for 
detected $\gamma$ rays with energy $\e_1$ 
gives a minimum Doppler factor
\begin{equation}
\dD^{min}= [ {\sigma_{\rm T} d_L^2 f_{\e_{pk}}\over 
4 m_ec^4 t_{v} \e_{pk}^A}\bigl({(1+z)^2 \e_1\over 2}
\bigr)^{1-A}\;]^{1/(6-2A)}\;
\label{delta}
\end{equation}
\citep{der05}, with $A =b$ if $\e_1 < \e_1^{pk}$, and $A = a$ if
$\e_1> \e_1^{pk}$. If $\dD^{min}\gtrsim \delta_{\phi\pi}$, we should
not expect GRBs to be neutrino bright.  Synchro-Compton analysis of
high-quality radio and gamma-ray blazar SED data with resolved VLBI
cores and self-absorption frequencies give the Doppler factor
directly, provided the synchrotron self-Compton component can be
identified and separated from external Compton and photohadronic
emission components. Should high-energy neutrinos be detectected when
the Doppler factor inferred from these tests is greater than
$\delta_{\phi\pi}$, then this would call into question our
understanding of the structure of black-hole jets, for example, the
assumption of isotropy of target photon distributions in the comoving
jet frame.

\section{Summary and Conclusions}

We have presented a detailed treatment of combined photomeson and
gamma-ray opacity, which is a crucial problem that unites the neutrino
particle physics and the electromagnetic worlds. This problem has been
numerically treated for GRBs \citep{da03}, but no detailed analytical
treatment is present in the scientific literature.  GLAST observations
will reveal if $\gamma$-ray spectra of FSRQ blazars and GRBs show
evidence for strong $\gamma$-ray absorption during periods of variable
target photon emissions, signifying favorable conditions for
high-energy neutrino production.

To illustrate the use of these results, suppose that a blazar or GRB with measured
redshift $z$ is monitored at optical, X-ray, or soft $\gamma$-ray energies, giving a light curve $f_\e(t)$
at photon energy $\e(t)$. The structure of the light curve implies $t_{var}(t)$.
From these observables, the Doppler factor $\delta_{\phi\pi}(t)$ for guaranteed importance of
photomeson production is derived from eq.\ (\ref{Epk1}). If high-energy 
neutrinos are detected, then the source must be opaque at $\gamma$-ray energies 
given by eq.\ (\ref{epk2}). If $\gamma$ rays are detected at these energies, 
then the basic relativistic jet model 
must be wrong. Suppose instead that $\gamma$ rays at some energy $\e_1$ are detected. In this case, 
the minimum Doppler factor $\dD^{min}(t)$can be inferred from eq.\ (\ref{delta})
to define times when these sources can and cannot be neutrino bright.

Times and locations of bright, variable MeV $\gamma$-ray and extincted
GeV fluxes from GRBs can be done exclusively with the GLAST GBM and
LAT, whereas other tests for $\gamma$-ray/multiwavelength correlations
giving the most favorable times for high-energy neutrino detection in
blazars require collaboration and coordination of separate facilities.
The necessary organization is already underway between GLAST and the
ground-based $\gamma$-ray telescopes, e.g., HESS and VERITAS, but
blazar observations with, e.g., Swift, Suzaku, and RXTE correlated
with GLAST, AGILE, and ground-based high-energy $\gamma$-ray
telescopes will be crucial for neutrino discovery science and testing
models of relativistic jet sources.

\acknowledgements 
We thank Armen Atoyan for discussions. The work of CD is supported by
the Office of Naval Research and a GLAST Interdisciplinary Scientist
Grant. TL's research is supported by the GLAST grant and a Swift Guest
Investigator Grant.

{}

\clearpage

\begin{figure}
%\vskip-2.5in
{\includegraphics[scale=0.6]{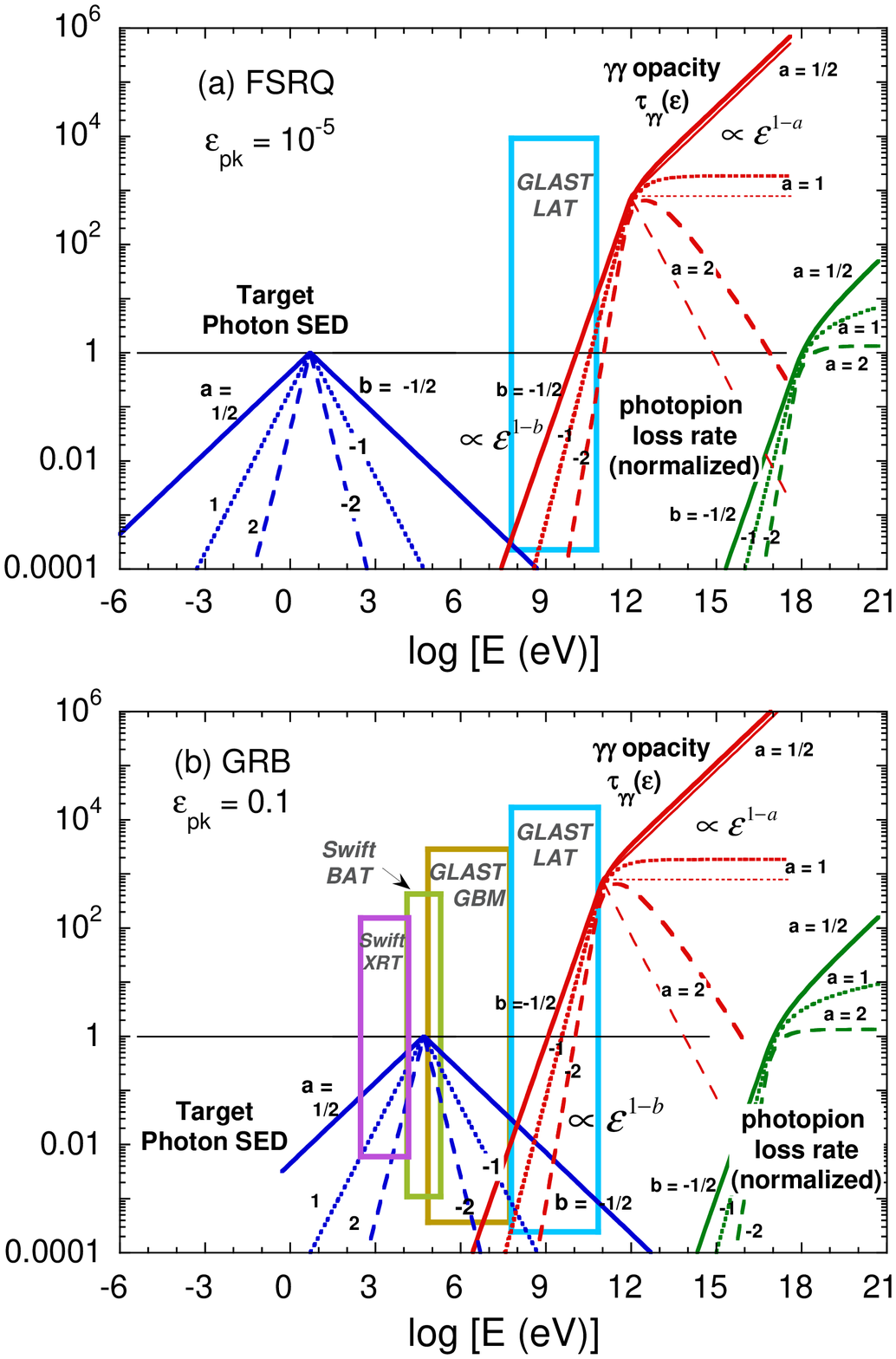}}
%\vskip-0.35in
\caption[]{
Target photon SED, $\gamma\gamma$ opacity $\tau_{\gamma\gamma}(\e_1)$,
and normalized photopion energy-loss rate $t^{\prime
-1}_{\phi\pi}/\rho_{\phi\pi}$ are shown as a function of photon or
neutrino energy for parameters of (a) optically flaring FSRQ and (b)
rapidly varying prompt GRB emission, using parameters from Table 1.
The target photon SED is approximated as a broken power law with $\nu
F_\nu$ flux peaking at $\e_{pk}$.  When photopion processes are
certain to be important for protons with energy $E_p^{\phi\pi}$ that
interact with peak target photons with energy $\approx \e_{pk}$, then
the $\gamma\gamma$ opacity of $\gamma$ rays with energy
$E_\gamma^{\gamma\gamma}$ is $\approx 800$. The $\gamma\gamma$ opacity
is less than unity at photon energies $\lesssim
E_\gamma^{\gamma\gamma}/800^{1/(1-b)}$. Heavy and light curves for
$\tau_{\gamma\gamma}(\e_1)$ are accurate numerical integrations
\citep{gs67} and $\delta$-function approximations \citep{der05},
respectively.}  \label{f1}
\end{figure} 

\clearpage

\begin{deluxetable}{lccccccc}
\tablecaption{Doppler Factor $\delta_{\phi\pi}$,
$\gamma$-Ray Photon Energy $E^{\gamma\gamma}_{\gamma} $, 
and Cosmic Ray Energy
$ E_p^{\phi\pi}$\label{table1}}
\tablehead{     &
\multicolumn{1}{c}{$\ell$\tablenotemark{a}} ~~& 
\multicolumn{1}{c}{$\eta$}&
$\tau$ & 
\multicolumn{1}{c}{$j$  }  &
\multicolumn{1}{c}{$\delta_{\phi\pi}$ } &
\multicolumn{1}{c}{$E_\gamma^{\gamma\gamma}({\rm GeV}) $  } &
\multicolumn{1}{c}{$ E_p^{\phi\pi}({\rm eV})$} 
}
\startdata
\tableline
\tableline
FSRQ& 28.7 & -11 & 5 &-5 (5 eV)& 9 & 92 & $5\times 10^{17}$ \\
$^{{\rm IR/optical}}$&  &  &  & -6 (0.5 eV)
& 16 & $30\times 10^3$ & $1.6\times 10^{19}$  \\
FSRQ& 28.7 & -11 & 5 &-2 (5 keV)& 1.6 & 0.03  & $1.6\times 10^{13}$  \\
$^{{\rm X-ray}}$&  &  &  & -3 (0.5 keV)
& 2.8 & 0.92 & $5\times 10^{14}$  \\
XBL & 27 & -10 & 3 &-2 (5 keV)& 1.3 & 0.14 & $3\times 10^{13}$  \\
$^{{\rm X-ray}}$&  &  &  & -3 (0.5 keV)
& 2.3 &4.7 & $9\times 10^{14}$  \\
GRB& 28.7 & -6 & 0 &0 (511 keV)& 160 & 2.9 & $ 2\times 10^{15}$  \\
$^{\gamma~{\rm ray}}$&  &  & &-1 (51 keV)& 280 & 92 & $5\times 10^{16}$  \\
$^{\rm X-ray~flare}$&  & -9 & 2&-3 (0.5 keV)& 50 & 290 & $1.6\times 10^{17}$ \\
\tableline
\enddata
%\end{tabular}
\tablenotetext{a}{Sources at $z = 2$ except for XBLs, at 
$z \approx 0.08$, $d_L = 10^{27}$ cm.}
\end{deluxetable}

\end{document}